\documentclass[final,3p,times,onecolumn]{elsarticle}
\onecolumn
\usepackage{hyperref}
\usepackage{pifont}
\usepackage{color}
\usepackage{tabularx, booktabs}
\usepackage{amsmath}
\usepackage{algorithmic}
\usepackage{float}
\usepackage{graphicx}
\usepackage{subcaption}

\usepackage{xcolor}
\usepackage{multirow}
\usepackage{algorithm}
\usepackage{hyperref}
\usepackage{caption}
\usepackage{fancyhdr}
\usepackage{tabularx, booktabs} 
\usepackage{natbib}
\usepackage{hyperref}
\usepackage{color}
\usepackage{tabularx, booktabs}
\usepackage{amsmath}
\usepackage{algorithmic}
\usepackage{float}

\usepackage{xcolor}
\usepackage{multirow}
\usepackage{algorithm}
\usepackage{hyperref}
\usepackage{caption}
\usepackage{fancyhdr}
\usepackage{tabularx, booktabs} 
\usepackage{natbib}
\usepackage{graphicx}
\newcommand{\cmark}{\ding{51}}%
\newcolumntype{C}[1]{>{\centering\arraybackslash}p{#1}}

\begin{document}

\begin{frontmatter}

\title{Toward Generalizable Cognitive Impairment Detection with Speech-Based Multimodal Large Language Models}

\author{Yingchao Huang\corref{mycorrespondingauthor}}
\ead{huangyi@saskpolytech.ca}
\address{Faculty of Digital Innovation, Arts \& Sciences, Saskatchewan Polytechnic, Regina SK S4S 5X1, Canada}
\author{Xin Wang}
\address{Faculty of Digital Innovation, Arts \& Sciences, Saskatchewan Polytechnic, Regina SK S4S 5X1, Canada}
\author{Yuhan Su}
\address{School of Basic Medical Sciences, Hebei University, Baoding 071000, China}
\author{Shanshan Yao}
\address{Department of Civil \& Environmental Engineering and School of Mining \& Petroleum Engineering, University of Alberta, Edmonton AB T6G 2H5, Canada}

\cortext[mycorrespondingauthor]{Corresponding author}


\begin{abstract}
Cognitive impairment (CI) is a growing public health concern. Early and accurate diagnosis is critical for enabling timely intervention and improving patient outcomes. Speech-based CI detection has emerged as a promising non-invasive approach, as speech signals encode both linguistic and acoustic markers associated with cognitive decline. Recent advances in large language models (LLMs) further strengthen the potential of speech-based assessment by enabling more expressive representation learning and improved generalization across diverse speakers, recording devices, and clinical environments. Moreover, multimodal learning by jointly modeling linguistic and acoustic features allows for a more comprehensive characterization of cognitive and behavioral changes related to CI, leading to more reliable detection. In this work, we propose a multimodal CI detection framework based on open-source LLMs that integrates speech audio and corresponding transcripts while preserving patient privacy. Acoustic embeddings are extracted directly from speech signals, while textual embeddings are generated from automatically transcribed speech. These modality-specific embeddings are then concatenated to create a combined feature vector and used for downstream classification, without requiring access to raw or sensitive patient data. The proposed approach is evaluated on the ADReSS20 and ADReSSo21 benchmark datasets. Experimental results show that the proposed multimodal framework achieves an CI classification accuracy of 92.4\% and consistently outperforms single-modality baselines. Our work establishes a new state-of-the-art for CI identification, with the proposed method demonstrating superior cross-dataset generalization. This advance highlights the power of an LLM-based multimodal framework that fuses linguistic and acoustic data to enable robust, scalable, and non-invasive screening.

\end{abstract}

\begin{keyword}
Cognitive Impairment, Speech-based diagnosis, Multimodal learning, Large language models, Acoustic and linguistic features, Privacy-preserving, Clinical generalization
\end{keyword} 
\end{frontmatter}

\section{Introduction}
\label{sec:intro}

Identifying cognitive impairment (CI) is critically important, as it often represents an early clinical manifestation of Alzheimer’s disease and other neurodegenerative conditions associated with progressive cognitive decline \cite{https://doi.org/10.1002/alz.13016}. Early detection of CI enables timely intervention, which can slow disease progression, improve patient outcomes, and enhance quality of life \cite{crous-bou2017alzheimer}. Early symptoms of CI typically include subtle changes in memory, language, and executive function, such as forgetfulness, word-finding difficulties, and impaired judgment \cite{Nowrangi2016NACC}. 

Among the various cognitive functions affected by CI, speech and language are often impaired at the earliest stages \cite{shahin2025zeroshotcognitiveimpairmentdetection}. Cognitive decline commonly manifests in speech through disfluencies, prolonged pauses, lexical simplification, and reduced semantic richness. Because these speech changes emerge gradually across different stages of disease progression, speech provides a sensitive and informative signal for monitoring cognitive trajectories and forecasting disease development \cite{ZOLNOORI2023102624}. Importantly, speech is a non-invasive and easily collectible biomarker that enables continuous and remote assessment, thereby reducing reliance on costly, time-consuming, and infrequent in-person clinical evaluations \cite{Ding2024SpeechSurvey}.

Speech-based CI detection is particularly appealing because it relies on natural, low-burden data that can be collected without specialized equipment. Moreover, speech is one of the most extensively studied modalities for cognitive assessment, supported by long-standing public datasets such as the Pitt Corpus \cite{becker1994natural}. The availability of these datasets has facilitated the development of a wide range of computational approaches for detecting CI across different disease stages. Consequently, automated speech-based CI detection has emerged as a promising solution for scalable, accessible, and cost-effective cognitive screening \cite{DeLaFuenteGarcia2020AIReview}.

Most existing speech-based CI detection methods rely on two primary categories of features: acoustic features and linguistic features. Acoustic features capture signal-level characteristics of speech, such as formants, pitch, and phonemes, and primarily reflect paralinguistic information related to speech production \cite{SHADLE2006442}. These features are often considered language-agnostic; however, variations in pronunciation, speaking style, and accent across languages and populations can still influence their reliability \cite{Martinc2020ADReSS, Edwards2020Multiscale, Luz2024Taukadial}.

In contrast, linguistic features focus on the content of speech and are derived from morphemes, words, phrases or sentences, and higher-level contextual meaning. These features are extracted either from manually annotated transcripts, when available \cite{Luz2020ADReSS}, or from transcripts generated by automatic speech recognition (ASR) systems \cite{OrtizPerez2024Multimodal, PerezToro2024Taukadial, ZHANG2025106821}. Linguistic features capture syntactic, semantic, and pragmatic properties of language, which are inherently language-dependent \cite{Zolnour2025LLMCARE}. While acoustic features characterize how speech is produced, linguistic features provide complementary information about what is being said, including contextual meaning and communicative intent \cite{shahin2025zeroshotcognitiveimpairmentdetection}.

Although prior studies have identified associations between specific acoustic markers and CI, acoustic changes alone do not reliably indicate pathological cognitive decline, particularly in older adults \cite{Badal2024AcousticCognitive}. Age-related physiological changes in speech production after the age of 60 can substantially alter acoustic patterns, making it difficult to disentangle normal aging effects from disease-related impairments \cite{MAHON202222, Galluzzi2018Presbyphonia, https://doi.org/10.1177/0194599815592373}. As a result, pre-trained deep learning models for acoustic feature extraction, such as wav2vec2.0 \cite{baevski2020wav2vec20frameworkselfsupervised} and VGGish \cite{Hershey2017CNNAudio}, may introduce bias when applied to elderly populations, as they are typically trained on large-scale datasets drawn from the general population. These limitations highlight the need to jointly model acoustic features alongside complementary modalities, while accounting for age-related variability in speech.

Motivated by these challenges, a growing body of research has explored multimodal approaches to improve the detection and longitudinal assessment of cognitive impairment \cite{POOR2024109199}. In general, integrating heterogeneous feature types within deep learning frameworks has been shown to enhance predictive performance \cite{10.1162/neco_a_01273, LAZLI2026109026, LU2024105669}. By combining multiple modalities, models can capture a broader set of indicators and exploit complementary information that may be more informative than any single modality in isolation \cite{HUANG2026108889, SONG2026108811}. Consequently, multimodal methods that fuse acoustic and linguistic features have gained increasing attention. Various fusion strategies including early fusion via feature concatenation, attention-based weighting, gating mechanisms, and late fusion of prediction scores have demonstrated improved robustness and accuracy by leveraging cross-modal complementarities \cite{Rohanian2020MultimodalAD, Wang2021ModularMultimodalAD, Koo2020MultimodalAD}. However, these multimodal systems also introduce additional challenges, including increased model complexity, feature alignment across modalities, temporal modeling, and generalization under dataset variability \cite{li2025multimodalalignmentfusionsurvey}.

Recent advances in large language models (LLMs) offer new opportunities to address several of these challenges. LLMs have demonstrated strong capability in capturing high-level linguistic patterns associated with cognitive decline and are increasingly used for CI detection by analyzing linguistic features such as grammar, word choice, and discourse structure \cite{Mo2024.08.22.24312463}. In typical pipelines, speech is first transcribed using ASR, after which LLMs generate text embeddings that encode semantic meaning and contextual usage. Extending beyond text-only analysis, Audio Large Language Models (AudioLLMs) enable joint processing of speech and text by integrating speech understanding capabilities into LLM architectures \cite{Peng2024SpeechLLMSurvey}. Representative AudioLLMs include WavLLM \cite{Hu2024WavLLM} and SALMONN \cite{Tang2024SALMONN}, which are based on LLaMA architectures, as well as Qwen-Audio \cite{Chu2023QwenAudio} and Qwen2-Audio \cite{Chu2024Qwen2Audio}, which are built upon the Qwen LLM family \cite{Bai2023Qwen}.

Despite their strong representational capacity, there is currently no established framework for effectively fusing acoustic and linguistic modalities within LLM-based systems, particularly when temporal characteristics of speech features are considered \cite{AlsuhaibaniMuath2025ARoM}. Furthermore, deploying LLM-based approaches in real-world healthcare environments presents significant practical challenges. Many state-of-the-art LLMs are closed-source, rely on external cloud-based services, and do not comply with healthcare data protection regulations such as Health Insurance Portability and Accountability Act (HIPAA), making them unsuitable for clinical deployment \cite{Price2019PrivacyBigData, Yadav2023DataPrivacyAI}. These constraints create a critical gap between methodological advances in LLM-based multimodal learning and their safe, privacy-preserving, and scalable adoption in clinical settings.

To bridge this gap, we propose a generalizable, multimodal, and privacy-preserving framework for CI detection from speech that integrates acoustic and linguistic representations using open-source LLMs and supports local deployment. Specifically, we leverage Qwen2-Audio to extract acoustic embeddings directly from speech and Qwen3 to generate linguistic embeddings from automatically transcribed text. These representations are fused in a shared latent space for CI classification. By jointly modeling how speech is produced and what is being said, the proposed framework aims to improve robustness, cross-dataset generalization, and clinical applicability for scalable cognitive impairment screening.

Compared with existing studies, the main contributions of this work are summarized as follows:

\begin{itemize}
\item Multimodal LLM-based pipeline: We propose a unified multimodal framework based on LLMs that integrates acoustic and linguistic speech representations. By fusing complementary modalities, the proposed pipeline effectively captures both the characteristics and content of speech, leading to improved CI detection performance.

\item Improved detection accuracy: The proposed framework achieves state-of-the-art performance on benchmark datasets, reaching an CI classification accuracy of 92.4\%. This result demonstrates a clear improvement over existing methods and highlights the effectiveness of multimodal LLM-based feature fusion.

\item Enhanced early CI detection: The proposed method demonstrates superior performance in identifying early-stage cognitive impairment compared with existing state-of-the-art models, enabling earlier clinical intervention and more proactive disease management.

\item Robust cross-dataset generalization: The framework is evaluated across multiple datasets and demonstrates strong generalization performance under dataset shifts, indicating its suitability for real-world clinical settings with diverse populations and recording conditions.

\item Privacy-preserving local deployment: Unlike prior approaches that rely on external cloud-based GPT APIs, the proposed framework is fully implemented using locally deployable open-source models, enabling compliance with data privacy regulations and safe deployment in privacy-sensitive clinical environments.
\end{itemize}

To support transparency and reproducibility, the implementation of the proposed framework and all experimental scripts will be made publicly available at https://github.com/kelci2017/CI\_Multimodal.

\section{Materials and methodology}
\label{methodology}

\subsection{Dataset and experiment setup}
The proposed multimodal framework is evaluated using two publicly available benchmark datasets, ADReSS20 \cite{luz2020alzheimersdementiarecognitionspontaneous} and ADReSSo21 \cite{luz2021detecting}, which are widely used for speech-based CI research. These datasets were developed as part of shared challenges and provide standardized evaluation protocols that facilitate reproducibility and fair comparison across studies. Both datasets include balanced diagnostic groups, Mini-Mental State Examination (MMSE) scores, and predefined training and testing splits stratified by age and gender, supporting both classification and regression analyses.

In both datasets, participants are recorded while performing a spontaneous speech task in which they describe the Cookie Theft picture from the Boston Diagnostic Aphasia Examination (BDAE) \cite{goodglass2001bdae}. This task is commonly employed in clinical and research settings to capture speech content, word usage, and sentence structure that are sensitive to cognitive decline. Alongside the speech recordings, MMSE scores are provided for cognitively normal (CN) participants and individuals diagnosed with CI. The BDAE and MMSE are established clinical tools routinely used to support cognitive assessment \cite{Teipel981}.

Although ADReSS20 and ADReSSo21 share the same speech task and diagnostic labels, they differ notably in data preparation and release formats, leading to meaningful dataset shifts. ADReSS20 includes full enhanced audio recordings as well as manually annotated transcripts and normalized, pre-segmented audio chunks. In contrast, ADReSSo21 provides only full enhanced audio recordings and does not include transcripts or segmented audio. As a result, the two datasets differ in temporal structure, segmentation granularity, and linguistic annotation availability, which can substantially influence models that rely on transcript-based features or fine-grained temporal modelling.

Beyond annotation differences, the datasets were collected from distinct speaker cohorts and recording sessions, introducing additional variability in speaker demographics, recording environments, microphone characteristics, and background noise. While both datasets are balanced by age and gender, these cohort- and environment-level differences introduce realistic covariate shifts in the speech signal. Such shifts pose a significant challenge for model generalization but closely reflect conditions encountered in real-world clinical deployment.

To minimize bias stemming from inconsistent annotation availability and to ensure a fair comparison across datasets, we retain only the full enhanced audio recordings from both ADReSS20 and ADReSSo21. All transcripts used in this study are generated automatically using a unified ASR pipeline. This design choice removes dependence on manual annotations while preserving real-world variability in speech quality and transcription noise.

In terms of dataset composition, ADReSS20 contains 54 CI and 54 CN participants in the training split, and 24 CI and 24 CN participants in the testing split. ADReSSo21 includes 87 CI and 79 CN participants for training, and 35 CI and 36 CN participants for testing. Based on these datasets, we define two complementary experimental configurations to assess both in-distribution performance and robustness under dataset shift.

In the first configuration, the training splits of ADReSS20 and ADReSSo21 are combined to form a larger training set, and their respective testing splits are similarly merged, while preserving the original age and gender stratification and keeping all test samples completely unseen during training. This combined setting results in 141 CI
and 133 CN participants in the training set and 59 CI and 60 CN participants in the testing set, yielding a total of 393 samples.

In the second configuration, cross-dataset generalization is explicitly evaluated by preserving the predefined training–testing splits of each dataset and training the proposed model on the training split of one dataset and directly testing it on the testing split of the other dataset, without any fine-tuning or overlap between datasets. This setting provides a rigorous evaluation of the ability of the framework to generalize across differences in cohort composition, recording conditions, preprocessing pipelines, and annotation availability.

By evaluating performance under both combined-dataset and cross-dataset scenarios, the experimental design enables a comprehensive assessment of the proposed framework’s robustness, generalization capability, and suitability for deployment in clinically realistic settings where data heterogeneity is unavoidable.

\subsection{Methodology}

This study proposed a generalizable, multimodal, and privacy-preserving framework for CI detection from spontaneous speech, leveraging the complementary strengths of AudioLLMs and large-scale text-based LLMs. The framework integrates acoustic embeddings extracted from patient speech and linguistic embeddings derived from automatically transcribed text, followed by feature normalization and classification using conventional machine learning and neural networks classifiers.

\begin{figure*}[!ht]
    \centering
    \includegraphics[width=0.9\textwidth]{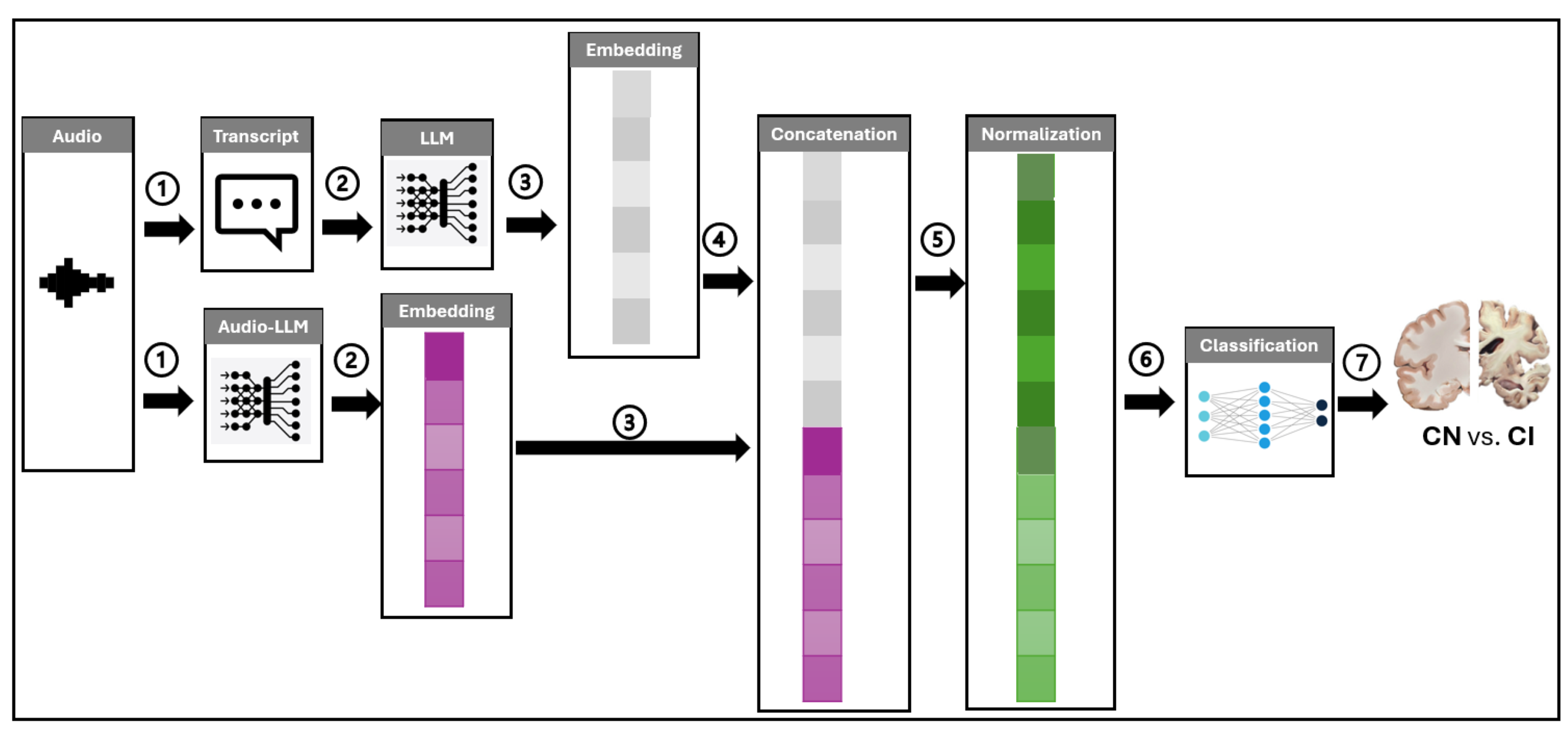}
    \caption{The workflow of the proposed framework.}
    \label{flow_fig}
\end{figure*}

As depicted in Figure~\ref{flow_fig}, the framework follows a parallel sequential processing pipeline as below:

Step \ding{172} Speech input and parallel processing: Speech audio recordings are used as the input. Each audio sample is processed in parallel through two pathways:
(i) ASR to generate textual transcripts, and
(ii) an AudioLLM to capture acoustic-level representations from the speech signal.

Step \ding{173} Modality-specific feature extraction: The ASR-generated transcripts are fed into a text-based LLM to extract high-level linguistic embeddings, while the speech audio is independently processed by the AudioLLM to generate audio embeddings that encode spectral and temporal speech characteristics.

Step \ding{174} Embedding generation: Both models output fixed-length, high-dimensional embedding vectors with text embeddings representing semantic, syntactic, and discourse-level language patterns, and audio embeddings representing acoustic and paralinguistic speech features associated with cognitive decline.

Step \ding{175} Multimodal feature fusion via concatenation: The extracted audio and text embeddings are concatenated along the feature dimension to form a unified multimodal representation, preserving complementary information from both speech content and speech production.

Step \ding{176} Embedding normalization: The concatenated multimodal embeddings are normalized to mitigate scale disparities between modalities, stabilize feature distributions, and improve optimization and generalization during classification.

Step \ding{177} Classification: The normalized multimodal embeddings are fed into a supervised classifier to learn discriminative patterns between cognitively normal and impaired speech.

Step \ding{178} Diagnostic prediction: The classifier outputs the final prediction, categorizing each subject’s speech sample as CN or CI.

\subsubsection{Audio Preprocessing}

The raw continuous-time speech signal is first subjected to preprocessing to reduce variability introduced by recording conditions, speaker characteristics, and dataset-specific acquisition protocols, and to ensure robust feature extraction for both ASR and audio embedding generation. This preprocessing stage standardizes the audio input across subjects and datasets, facilitating consistent linguistic transcription and stable acoustic representation learning. Specifically, a sequence of operations including resampling, amplitude normalization, non-speech removal, and segmentation, is applied to mitigate noise, align temporal and spectral properties, and improve the reliability and comparability of downstream ASR outputs and AudioLLM-derived embeddings.

Given a raw speech signal \( x_i \), the signal is initially resampled with the resampling operation  $\mathcal{R}(\cdot)$ to a fixed sampling rate \( f_s \) to ensure temporal consistency across recordings:

\begin{equation}
x_i^{(1)} = \mathcal{R}_{f_s}(x_i^{\text{raw}}),
\end{equation}

If the recording contains multiple channels, it is converted to a mono signal by channel averaging:
\begin{equation}
x_i^{(2)} = \frac{1}{C} \sum_{c=1}^{C} x_{i,c}^{(1)},
\end{equation}
where $C$ is the number of channels. This step eliminates channel-dependent variability and ensures a uniform waveform representation.

To reduce variations caused by microphone gain and recording distance, the waveform is amplitude-normalized:
\begin{equation}
x_i^{(3)} = \frac{x_i^{(2)}}{\max\left(|x_i^{(2)}|\right)}.
\end{equation}
This normalization stabilizes the dynamic range of the signal and improves robustness for both ASR and acoustic embedding extraction.

Non-speech regions such as silence and background noise are removed using voice activity detection:
\begin{equation}
x_i^{(4)} = \mathcal{V}(x_i^{(3)}),
\end{equation}
where $\mathcal{V}(\cdot)$ denotes a speech activity filtering operator. This step focuses subsequent processing on linguistically and acoustically informative regions.

The filtered speech signal is segmented into fixed-length chunks of $L$ seconds:
\begin{equation}
x_i^{(5)} = \{ x_{i,j}^{(4)} \}_{j=1}^{M_i}, \quad x_{i,j}^{(4)} \in \mathbb{R}^{f_s \cdot L},
\end{equation}
where $M_i$ is the number of segments for subject $i$. Segmenting long recordings improves memory efficiency and enables temporal aggregation of embeddings.

\subsubsection{Acoustic Embedding Extraction}

Each speech segment is encoded into a high-level acoustic representation using an AudioLLM encoder:
\begin{equation}
\mathbf{a}_{i,j} = \Phi_{\text{audio}}(x_{i,j}^{(5)}),
\end{equation}
where $\Phi_{\text{audio}}(\cdot)$ maps raw audio to a fixed-dimensional embedding that captures prosodic, spectral, and temporal speech characteristics.

Segment-level embeddings are aggregated using temporal average pooling to obtain a subject-level acoustic embedding:
\begin{equation}
\mathbf{a}_i = \frac{1}{M_i} \sum_{j=1}^{M_i} \mathbf{a}_{i,j}.
\end{equation}
This aggregation yields a compact representation while preserving global acoustic patterns across the recording.

\subsubsection{Automatic Speech Recognition and Text Processing}

Each audio segment is transcribed using an ASR model:
\begin{equation}
\hat{t}_{i,j} = \Psi_{\text{ASR}}(x_{i,j}^{(5)}),
\end{equation}
where $\Psi_{\text{ASR}}(\cdot)$ denotes the transcription function.

Segment-level transcripts are concatenated to form the full subject transcript:
\begin{equation}
\hat{t}_i = \bigcup_{j=1}^{M_i} \hat{t}_{i,j}.
\end{equation}

The transcript is normalized to remove non-linguistic symbols and standardize formatting:
\begin{equation}
\hat{t}_i^{(1)} = \mathcal{N}(\hat{t}_i),
\end{equation}
followed by tokenization:
\begin{equation}
\hat{t}_i^{(2)} = \mathcal{T}(\hat{t}_i^{(1)}),
\end{equation}
where $\mathcal{N}(\cdot)$ and $\mathcal{T}(\cdot)$ denote text normalization and tokenization operators. These steps ensure consistency for language model encoding.

\subsubsection{Linguistic Embedding Extraction}

The processed transcript is encoded into a linguistic embedding using a language LLM:
\begin{equation}
\mathbf{l}_i = \Phi_{\text{text}}(\hat{t}_i^{(2)}),
\end{equation}
where $\Phi_{\text{text}}(\cdot)$ produces a high-level semantic representation capturing speech content, word usage, and sentence structure associated with cognitive status.

\subsubsection{Multimodal Feature Fusion and Normalization}

Acoustic and linguistic embeddings are concatenated to form a multimodal representation:
\begin{equation}
\mathbf{z}_i = [\mathbf{a}_i \; \| \; \mathbf{l}_i].
\end{equation}
This fusion preserves complementary information from speech production and speech content.

To mitigate scale disparities between modalities and stabilize classifier training, Min--Max normalization is applied:
\begin{equation}
\tilde{\mathbf{z}}_i =
\frac{\mathbf{z}_i - \mathbf{z}_{\min}}
{\mathbf{z}_{\max} - \mathbf{z}_{\min}},
\end{equation}
where $\mathbf{z}_{\min}$ and $\mathbf{z}_{\max}$ are computed from the training set.

\subsubsection{Classification}

The normalized multimodal embedding is fed into a supervised classifier to predict the cognitive label:
\begin{equation}
\hat{y}_i = f(\tilde{\mathbf{z}}_i),
\end{equation}
where $f(\cdot)$ represents a classification model, including support vector classifier (SVC), logistic regression, XGBoost, or neural networks.

\section{Results and discussion}

To comprehensively assess the effectiveness of the proposed multimodal framework, this section provides a thorough evaluation of the proposed method in terms of classification performance, robustness, and generalization across heterogeneous datasets. We conducted extensive experiments on the ADReSS20 and ADReSSo21 benchmark datasets and systematically compared its performance with representative state-of-the-art methods reported in the literature. In addition to standard performance evaluation, ablation analyses were conducted to quantify the contribution of individual components within the proposed pipeline, and cross-dataset generalization studies were performed to evaluate resilience to dataset shifts. 

\subsection{Testing results discussion}

Motivated by prior studies demonstrating the effectiveness of embedding-based classifiers for CI detection \cite{SHARMA2021100012, eng6070163, Mortensen2025ADetectoLocum}, where embeddings extracted from LLMs are commonly used as input features for downstream classification \cite{Kashyap2025ExplainableDementiaLLM, eng6070163, Mo2024.08.22.24312463}, we evaluate a set of widely adopted downstream classifiers that have shown strong performance when operating on high-level embedding representations. Specifically, we consider both classical machine learning models and shallow neural networks, and optimize their hyperparameters using grid search \cite{liashchynskyi2019gridsearchrandomsearch} within a five-fold cross-validation framework. Following this established evaluation paradigm, logistic regression, SVC, XGBoost, and neural networks are selected as downstream classifiers to enable direct and meaningful comparison with existing approaches in the literature \cite{XIAO2021102362, SHANMUGAM2022103217, BOTROS2025106997}. This unified training and validation strategy ensures a fair and systematic comparison across classifiers under consistent experimental conditions.

Tables~\ref{cv_table} and~\ref{test_table} summarize the cross-validation and held-out testing results, respectively, providing insights into both training stability and generalization to unseen data. During cross-validation (Table~\ref{cv_table}), classical machine learning models including SVC, XGBoost, and logistic regression exhibit relatively strong and stable performance, achieving accuracies in the range of 82-85\% with comparatively low standard deviations across folds. In contrast, the neural networks classifier shows lower average cross-validation performance and higher variability, particularly in recall and F1-score. This behavior is consistent with expectations in small-to-moderate sample size settings, where neural networks are more sensitive to data partitioning and may struggle to reliably estimate a large number of parameters. Similar observations have been reported in prior work, which shows that neural networks can underperform linear or kernel-based models during cross-validation when training data are limited or highly heterogeneous \cite{goodfellow2016deep, hastie2009elements}.

However, a markedly different trend emerges in the final testing results (Table \ref{test_table}). When evaluated on the unseen test set, the neural networks substantially outperforms all other classifiers, achieving an accuracy of 92.4\% and an F1-score of 92.2\%, exceeding the best-performing classical models by a clear margin. This performance gap indicates that the neural networks are better able to leverage the full representational capacity of the multimodal embeddings once trained on the complete training data, despite its higher variance during cross-validation.

A key factor contributing to this improvement is the high dimensionality and heterogeneity of the concatenated multimodal embeddings. By fusing audio and text representations, the feature space grows substantially, capturing complementary acoustic and linguistic cues relevant to CI detection. Neural networks are particularly well-suited for such settings, as they can model non-linear interactions across modalities and automatically learn hierarchical feature combinations that are difficult to capture with shallow models \cite{lecun2015deep, bengio2013representation}.

In contrast, logistic regression is a linear classifier that assumes a linear decision boundary in the input space. While it performs competitively in terms of precision, its limited capacity restricts its ability to fully exploit complex cross-modal relationships in high-dimensional embeddings, leading to lower recall and F1-score on the test set. This limitation of linear models in high-dimensional, multimodal contexts is well documented \cite{hastie2009elements}.

Similarly, although XGBoost is a powerful ensemble method, tree-based models often struggle with very high-dimensional dense feature representations, especially when features are highly correlated, as is typical for concatenated neural embeddings \cite{chen2016xgboost}. Such conditions can reduce the effectiveness of tree splits and limit generalization performance. SVC, while robust in moderate-dimensional spaces, can also suffer from scalability and representational limitations when the feature dimension becomes large relative to the number of training samples \cite{cortes1995support}.

\begin{table}[]
\caption{Model training results with standard deviations using cross-validation. }
\begin{center}
\begin{tabular}{C{2cm}|C{1.2cm}|C{1.2cm}|C{1.2cm}|C{1.2cm}|C{1.5cm}|C{1.5cm}|C{1.5cm}|C{1.5cm}}
\hline
\textbf{Classifier}          & \textbf{Accuracy} & \textbf{Precision} & \textbf{Recall} & \textbf{F1} & \textbf{cv accuracy std} & \textbf{cv precision std} & \textbf{cv recall std} & \textbf{cv f1 score std} \\ \hline
Neural Networks     & 77.0\%            & 56.0\%             & 46.7\%          & 50.5\%      & 6.0\%                    & 12.1\%                    & 17.4\%                 & 14.8\%                   \\ \hline
SVC                & 84.7\%            & 86.2\%             & 84.4\%          & 85.0\%      & 4.8\%                    & 6.7\%                     & 6.2\%                  & 4.5\%                    \\ \hline
XGBoost             & 82.1\%            & 83.6\%             & 81.6\%          & 82.4\%      & 2.6\%                    & 4.6\%                     & 4.7\%                  & 2.6\%                    \\ \hline
Logistic Regression & 84.3\%            & 87.4\%             & 81.6\%          & 84.3\%      & 5.9\%                    & 7.4\%                     & 6.2\%                  & 5.9\%                    \\ \hline
\end{tabular}
\end{center}
\label{cv_table}
\end{table}

\begin{table}[]
\caption{Model testing results. }
\begin{center}
\begin{tabular}{c|c|c|c|c}
\hline
\textbf{Classifier}          & \textbf{Accuracy} & \textbf{Precision} & \textbf{Recall} & \textbf{F1}     \\ \hline
Neural Networks     & \textbf{92.4\%}   & \textbf{94.6\%}    & \textbf{89.8\%} & \textbf{92.2\%} \\ \hline
SVC                & 89.9\%            & 91.2\%             & 88.1\%          & 89.7\%          \\ \hline
XGBoost             & 84.9\%            & 87.3\%             & 81.4\%          & 84.2\%          \\ \hline
Logistic Regression & 89.9\%            & 94.3\%             & 84.7\%          & 89.3\%          \\ \hline
\end{tabular}
\end{center}
\label{test_table}
\end{table}

\subsection{Early detection analysis}
\begin{figure}[!ht]
    \centering
    \includegraphics[width=0.7\textwidth]{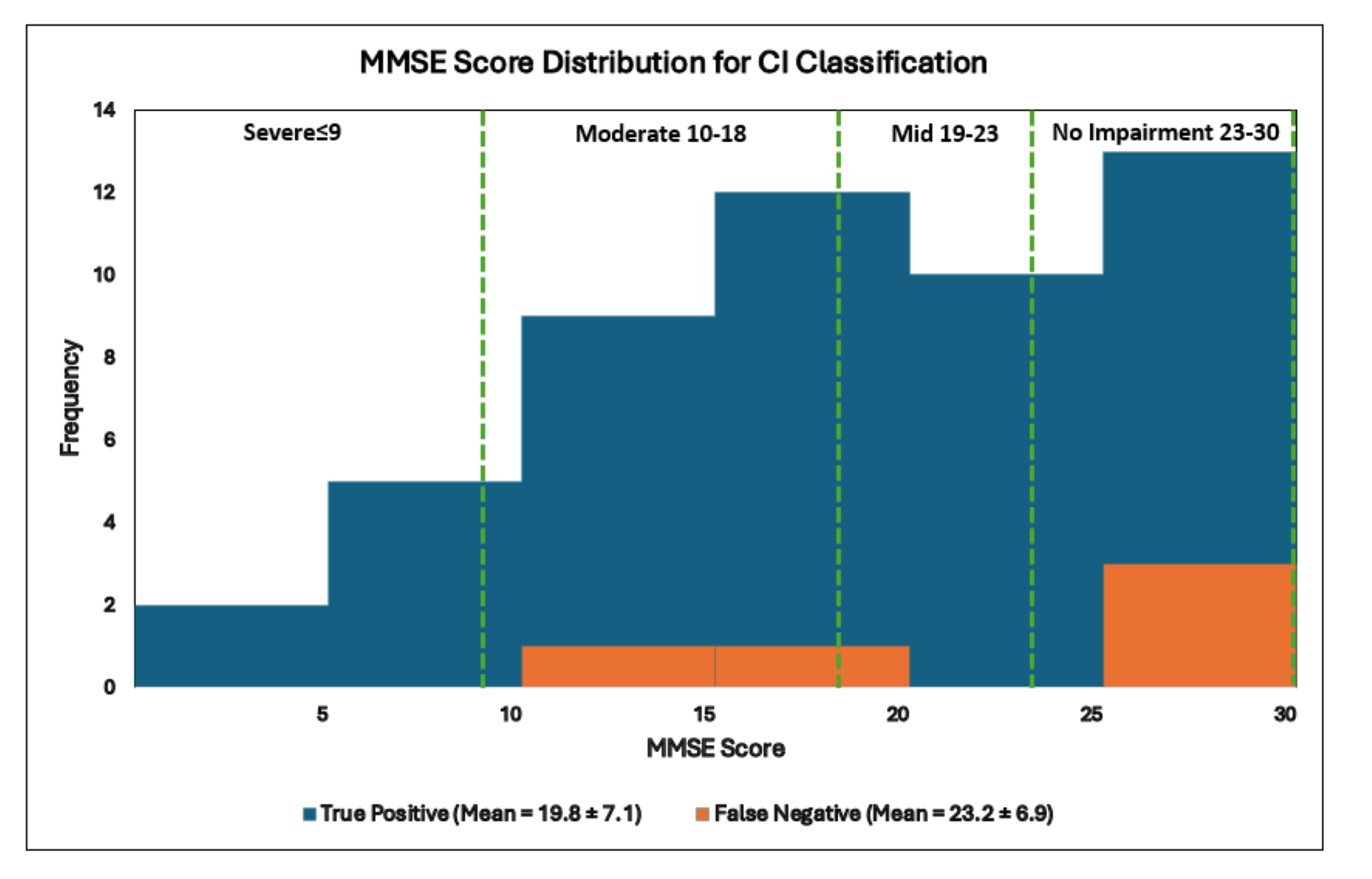}
    \caption{Histogram of correct and incorrect CI classifications with thresholds indicating MMSE cognitive impairment level. Means and standard deviations are calculated for MMSE scores per class.}
    \label{mmse_fig}
\end{figure}

A central clinical objective of the proposed framework is the reliable identification of CI, particularly during the early stages of cognitive decline when timely intervention can have the greatest impact. Figure~\ref{mmse_fig} evaluates this capability by analyzing model classification outcomes in relation to MMSE scores, a widely used cognitive screening tool in clinical practice. In the MMSE, higher scores indicate preserved cognitive function, while lower scores reflect increasing levels of impairment \cite{Kurlowicz1999MMSE}.

Figure~\ref{mmse_fig} presents the distribution of MMSE scores for CI cases correctly identified by the proposed multimodal framework (true positives) and those misclassified as cognitively normal (false negatives). The MMSE ranges are annotated according to standard clinical thresholds: severe impairment ($\leq$9), moderate impairment (10--18), mild impairment (19--23), and no cognitive impairment (24--30).

The majority of true positive CI cases are concentrated within the moderate (10--18) and mild (19--23) impairment ranges, with a mean MMSE score of $19.8 \pm 7.1$. This distribution demonstrates that the proposed framework effectively detects CI across clinically relevant stages of cognitive decline, including early and transitional phases where diagnostic uncertainty is highest. Notably, a substantial proportion of correctly identified cases fall within the mild impairment range, highlighting the model’s sensitivity to subtle cognitive changes that are often difficult to capture using unimodal approaches.

This strong performance in the mild CI range underscores the advantage of multimodal feature integration, as combining acoustic and linguistic representations enables the model to capture early pathological speech patterns before severe cognitive deterioration becomes apparent. Such capability is particularly valuable for early screening and proactive clinical management.

In contrast, false negative cases are sparsely distributed across the moderate, mild, and higher MMSE ranges, with a mean score of $23.2 \pm 6.9$. These misclassified subjects are concentrated near the upper end of the MMSE spectrum. These misclassified subjects are located toward the upper end of the MMSE spectrum. This pattern indicates that misclassifications primarily occur near the clinical decision boundary between normal cognition and early cognitive decline, where impairments are subtle and may not be consistently reflected by global screening measures. Such cases highlight the inherent challenge of detecting very early CI using behavioural assessments alone and further underscore the importance of sensitive multimodal biomarkers with the combined acoustic and linguistic representations leveraged in the proposed framework for improving early-stage detection accuracy.


\subsection{Comparative results discussion}
To provide a comprehensive evaluation of the proposed framework, we compare its performance with several representative methods reported in the literature that were developed and evaluated on the ADReSS20 and ADReSSo21 datasets. This comparison situates the proposed approach within the broader landscape of speech-based CI classification and enables a direct assessment of its effectiveness relative to established state-of-the-art techniques.

Table~\ref{comp_table} summarizes the comparative results. Overall, the proposed multimodal framework consistently outperforms prior approaches across all evaluation metrics, demonstrating clear and substantial performance gains. Unlike earlier methods that rely primarily on single-modal features or loosely coupled architectures, our approach integrates acoustic and linguistic embeddings within a unified multimodal learning framework, resulting in more discriminative and robust representations for CI detection.

Specifically, our method achieves the highest overall accuracy of 92.4\%, surpassing the strongest reported baseline by Mortensen et al. \cite{mortensen2025early} by 7.5 percentage points. This improvement is particularly noteworthy given that they employ a strong ensemble-style pipeline that combines a LLM (Zephyr) with gradient-boosted trees. The superior performance of our approach therefore reflects the effectiveness of multimodal representation fusion rather than increased model complexity or classifier strength alone.

In addition to accuracy, the proposed framework demonstrates a marked advantage in precision, achieving 94.6\%, which is substantially higher than all compared methods. High precision is especially important in clinical screening scenarios, where false positive classifications can lead to unnecessary anxiety, additional testing, and increased clinical burden. At the same time, our model maintains a strong recall of 89.8\%, indicating high sensitivity to CI cases and ensuring that improved specificity is not achieved at the expense of missed diagnoses. The resulting F1-score of 92.2\% reflects a well-balanced trade-off between precision and recall and clearly exceeds the performance of all prior approaches.

In contrast, several earlier methods exhibit notable trade-offs. For example, Agbavor et al.~\cite{Agbavor2023AIADVoice} report very high recall (97.1\%) but substantially lower precision (72.3\%), suggesting a tendency to over-detect CI cases. Other approaches based on traditional classifiers or text-only representations achieve more balanced performance but remain consistently below the proposed framework across all metrics. Collectively, these comparisons highlight that the proposed multimodal framework not only improves overall classification accuracy but also delivers a clinically meaningful balance between sensitivity and specificity, underscoring its suitability for real-world cognitive impairment screening.

\begin{table}[]
\caption{Comparative results of our model results with other contributors for CI identification. }
\begin{center}
\begin{tabular}{c|c|c|c|c|c}
\hline
\textbf{Contributors} & \textbf{Model}   & \textbf{Accuracy} & \textbf{Precision} & \textbf{Recall} & \textbf{F1}    \\ \hline
Mortensen et al. \cite{mortensen2025early} & Zephyr + XGB     & 84.9\%             & 84.7\%              & 84.7\%           & 84.7\%          \\ \hline
Bang et al. \cite{https://doi.org/10.4218/etrij.2023-0356}      & BERT             & 83.1\%             & 83.1\%              & 83.1\%           & 83.1\%          \\ \hline
Agbavor et al. \cite{Agbavor2023AIADVoice}   & GPT-3.5 + SVC    & 80.3\%             & 72.3\%              & 97.1\%           & 82.9\%          \\ \hline
Luz et al. \cite{luz2021detecting}       & SVC              & 78.9\%             & 77.8\%              & 80.0\%           & 78.9\%          \\ \hline
\textbf{Ours}         & \textbf{Multimodal (Qwen-based)} & \textbf{92.4\%}    & \textbf{94.6\%}     & \textbf{89.8\%}  & \textbf{92.2\%} \\ \hline
\end{tabular}
\end{center}
\label{comp_table}
\end{table}

The proposed framework leverages Qwen-based models for both acoustic and linguistic representation learning, employing different model variants to better align model capacity with modality-specific requirements. Specifically, the AudioLLM component utilizes Qwen2.5-7B for extracting acoustic embeddings from speech signals, while textual embeddings are generated using Qwen3-30B. This asymmetric design is intentional and reflects the differing representational demands of audio and text modalities, rather than a uniform choice of model size across components.

While Qwen2.5-7B provides sufficient capacity for extracting high-level acoustic representations from speech, where temporal and spectral patterns are relatively constrained, the generation of text embeddings benefits more substantially from increased model scale. Linguistic representations require deeper contextual reasoning to capture subtle semantic, syntactic, and discourse-level patterns that are particularly relevant for CI detection. For this reason, we employ Qwen3-30B for transcript encoding rather than using the same 7B-scale model for both modalities.

To empirically validate this design choice, we conducted a controlled comparison in which Qwen2.5-7B and Qwen3-30B were alternatively used as transcript encoders while keeping all other components fixed, including the AudioLLM embeddings, feature normalization strategy, and classification pipelines. The results, reported in Table~\ref{awen_table}, isolate the effect of text representation capacity on overall multimodal performance.

Across all classifiers, Qwen3-30B consistently outperforms Qwen2.5-7B, with the most pronounced gains observed when paired with the neural networks classifier. In particular, the Qwen3-30B + neural networks configuration achieves an accuracy of 92.4\% and an F1 score of 92.2\%, representing an improvement of approximately five percentage points over the corresponding Qwen2.5-7B configuration. Given the balanced nature of the datasets and the fixed experimental setup, these improvements reflect a meaningful enhancement in discriminative capability rather than marginal optimization effects.

The superior performance of Qwen3-30B can be attributed to its larger parameter capacity and enhanced contextual modeling ability, which enable more accurate encoding of fine-grained linguistic cues associated with early cognitive impairment. These cues including reduced semantic specificity, syntactic simplification, and diminished discourse coherence are often subtle and distributed across long speech segments, making them particularly well-suited to large-scale language models with stronger contextual reasoning. When combined with acoustic embeddings, the richer textual representations produced by Qwen3-30B provide complementary information that can be more effectively exploited by expressive classifiers.

Notably, classical classifiers such as SVC and logistic regression exhibit smaller performance differences between Qwen2.5-7B and Qwen3-30B. This observation is consistent with their limited ability to model complex nonlinear interactions and high-dimensional feature spaces. In contrast, the neural networks classifier benefits most from the higher-dimensional and more expressive text embeddings generated by Qwen3-30B, underscoring the importance of matching representation capacity with classifier expressiveness in multimodal learning settings.

\begin{table}[]
\caption{The test performance comparisons with different Qwen-based model sizes using the same setups. }
\begin{center}
\begin{tabular}{c|c|c|c|c|c}
\hline
\textbf{Model}                      & \textbf{Classifier} & \textbf{Accuracy} & \textbf{Precision} & \textbf{Recall} & \textbf{F1} \\ \hline
\multirow{4}{*}{\textbf{Qwen2.5-7B}}  & Neural Networks      & 87.4\%            & 87.9\%             & 86.4\%          & 87.2\%      \\ \cline{2-6} 
                                    & SVC                 & 89.9\%            & 89.8\%             & 89.8\%          & 89.8\%      \\ \cline{2-6} 
                                    & XGBoost             & 84.9\%            & 84.7\%             & 84.7\%          & 84.7\%      \\ \cline{2-6} 
                                    & Logistic Regression & 89.1\%            & 88.3\%             & 89.8\%          & 89.1\%      \\ \hline
\multirow{4}{*}{\textbf{Qwen3-30B}} & Neural Networks      & 92.4\%            & 94.6\%             & 89.8\%          & 92.2\%      \\ \cline{2-6} 
                                    & SVC                 & 89.9\%            & 91.2\%             & 88.1\%          & 89.7\%      \\ \cline{2-6} 
                                    & XGBoost             & 84.9\%            & 87.3\%             & 81.4\%          & 84.2\%      \\ \cline{2-6} 
                                    & Logistic Regression & 89.9\%            & 94.3\%             & 84.7\%          & 89.3\%      \\ \hline
\end{tabular}
\end{center}
\label{awen_table}
\end{table}

\subsection{Ablation study}

To quantify the contribution of each component in the proposed multimodal framework, we conduct a comprehensive ablation study by selectively removing audio embeddings, text embeddings, and the normalization step applied to the concatenated feature representations. The results, summarized in Table~\ref{abl_table}, clearly demonstrate that each component plays a critical and complementary role in achieving optimal performance.

When only audio embeddings are used with normalization (first row), the model achieves an accuracy of 82.4\% and an F1 score of 81.4\%, indicating that acoustic cues alone provide meaningful information for CI identification. However, removing normalization in the audio-only setting (second row) leads to a substantial performance degradation (74.8\% accuracy and 74.1\% F1), highlighting the importance of feature scaling even in unimodal acoustic representations to stabilize training and reduce sensitivity to amplitude variability.

A similar pattern is observed for text-only configurations. Without normalization (third row), the text-only model attains relatively low accuracy (68.9\%) despite a high recall (88.1\%), suggesting that linguistic embeddings are highly sensitive to CI-related patterns but may produce unstable or biased predictions when feature distributions are not properly aligned. Applying normalization to text-only embeddings (fourth row) yields a marked improvement, achieving 85.7\% accuracy and an 85.5\% F1 score. This result demonstrates that normalization is particularly critical for high-dimensional linguistic embeddings to control variance and enable effective downstream classification.

When both audio and text embeddings are combined but normalization is omitted (fifth row), performance drops sharply (74.8\% accuracy and 73.7\% F1), despite the availability of richer multimodal information. This outcome underscores that simply concatenating heterogeneous embeddings without proper normalization can lead to feature imbalance, where one modality dominates the learning process and degrades overall performance.

The best performance is achieved when all three components including audio embeddings, text embeddings, and normalizationare jointly employed (last row), resulting in an accuracy of 92.4\% and an F1 score of 92.2\%. Compared to the strongest unimodal configurations, the full framework provides a substantial performance gain, confirming that acoustic and linguistic embeddings capture complementary aspects of cognitive impairment. The normalization step further ensures that these heterogeneous representations are aligned within a shared feature space, enabling the classifier to effectively exploit cross-modal interactions.

\begin{figure}[!ht]
    \centering
    \includegraphics[width=0.7\textwidth]{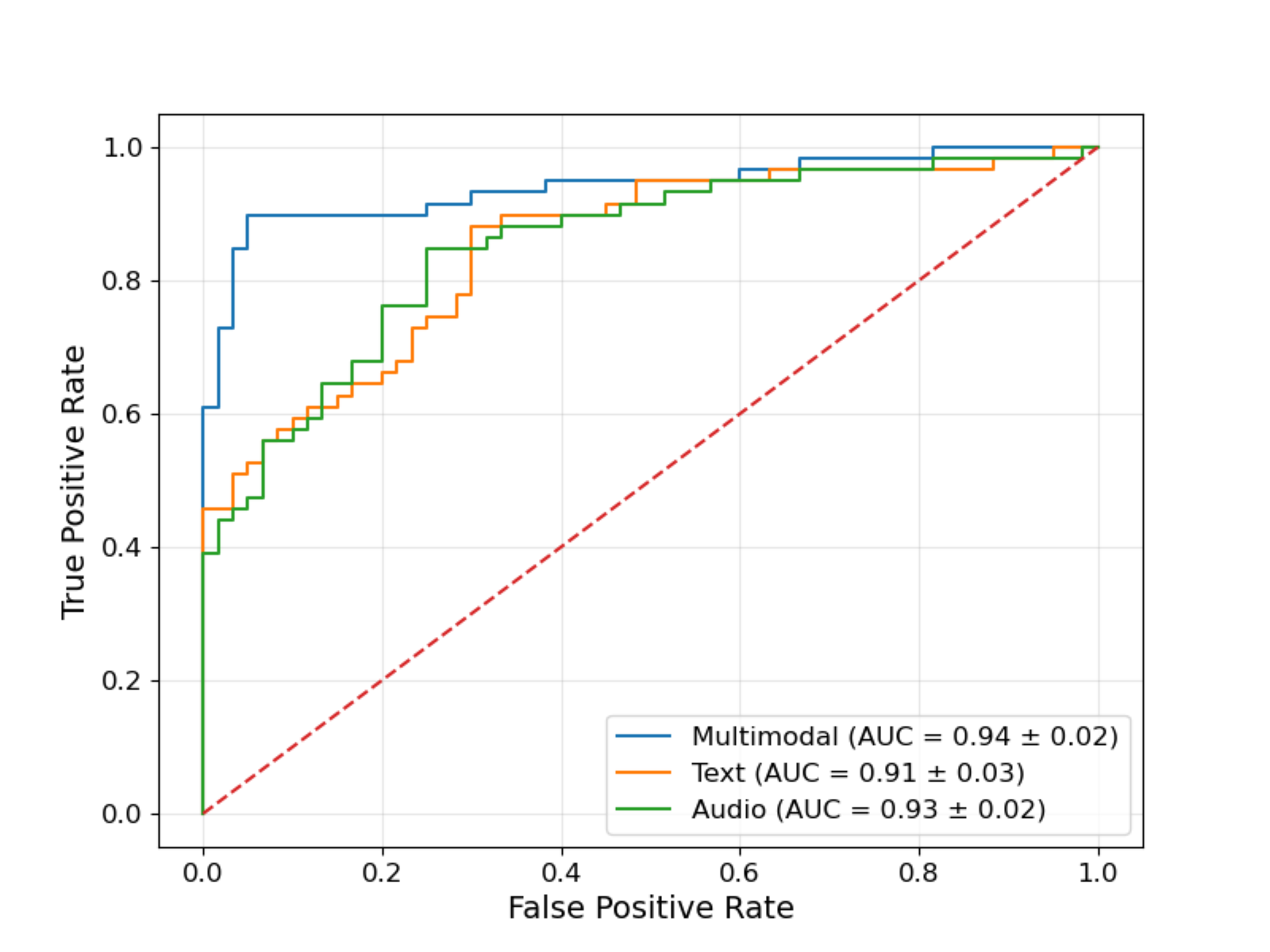}
    \caption{ROC curve comparing multimodal and single modalities with 5-fold cross validation derived AUC standard deviations.}
    \label{roc_fig}
\end{figure}

To further assess the discriminative contribution of multimodal fusion beyond point-based metrics, we perform receiver operating characteristic (ROC) analysis. Figure~\ref{roc_fig} compares the ROC curves of the proposed multimodal framework with its audio-only and text-only counterparts. The ROC curve illustrates the trade-off between the true positive rate (TPR) and false positive rate (FPR) across decision thresholds, while the area under the curve (AUC) provides a threshold-independent measure of classification performance.

The multimodal model achieves the highest overall performance, with an AUC of $0.94 \pm 0.02$, outperforming both the audio-only model ($0.93 \pm 0.02$) and the text-only model ($0.91 \pm 0.03$). This consistent improvement confirms that integrating acoustic and linguistic representations enhances discriminative capability beyond either modality alone. Notably, the multimodal ROC curve dominates the single-modality curves across most FPR regions, particularly in the low-false-positive regime, which is clinically critical for minimizing false alarms in screening scenarios.

While the audio-only model performs competitively, reflecting the diagnostic value of prosodic and fluency-related cues, and the text-only model demonstrates strong sensitivity to linguistic markers of CI, both exhibit limitations when used in isolation. The superior ROC performance of the multimodal framework indicates that acoustic and linguistic features provide complementary information. By fusing normalized embeddings from both modalities, the proposed framework achieves more robust, balanced, and generalizable CI detection.

\begin{table}[]
\caption{Ablation study to reflect the contribution of each component in the proposed method. }
\begin{center}
\begin{tabular}{c|c|c|c|c|c|c}
\hline
\textbf{Audio   Embeddings} & \textbf{Text Embeddings} & \textbf{Normalization} & \textbf{Accuracy} & \textbf{Precision} & \textbf{Recall} & \textbf{F1} \\ \hline
\cmark                      & \ding{55}                & \cmark                 & 82.4\%            & 85.2\%             & 78.0\%          & 81.4\%      \\ \hline
\cmark                      & \ding{55}                & \ding{55}              & 74.8\%            & 75.4\%             & 72.9\%          & 74.1\%      \\ \hline
\ding{55}                   & \cmark                   & \ding{55}              & 68.9\%            & 63.4\%             & 88.1\%          & 73.8\%      \\ \hline
\ding{55}                   & \cmark                   & \cmark                 & 85.7\%            & 86.2\%             & 84.7\%          & 85.5\%      \\ \hline
\cmark                      & \cmark                   & \ding{55}              & 74.8\%            & 76.4\%             & 71.2\%          & 73.7\%      \\ \hline
\cmark                      & \cmark                   & \cmark                 & 92.4\%            & 94.6\%             & 89.8\%          & 92.2\%      \\ \hline
\end{tabular}
\end{center}
\label{abl_table}
\end{table}

\subsection{Model generalization discussion}

To rigorously evaluate the generalization capability of the proposed multimodal framework, we conduct cross-dataset validation experiments in which the model is trained on one dataset and directly tested on the other, without any fine-tuning or adaptation. This evaluation setting reflects a clinically realistic deployment scenario, where models must operate robustly under dataset shifts arising from differences in cohort composition, recording conditions, and preprocessing pipelines.

Training on ADReSS20 and testing on ADReSSo21 (Table \ref{gen_table1}) demonstrates strong generalization performance for the proposed method. The neural networks classifier achieves an accuracy of 91.5\% with an F1 score of 91.2\%, indicating that the learned multimodal representations transfer effectively despite variability in speaker demographics, recording environments and microphone characteristics between the two datasets. XGBoost also shows competitive performance (88.7\% accuracy), whereas SVC and logistic regression exhibit substantial degradation. This performance gap suggests that classifiers with limited capacity to model nonlinear interactions struggle to adapt to cross-dataset distribution shifts, particularly when high-dimensional multimodal embeddings are involved.

Training on ADReSSo21 and testing on ADReSS20 (Table \ref{gen_table2}) yields similarly strong results, further confirming the robustness of the proposed framework. The neural networks again achieves 91.7\% accuracy and F1 score, matching its in-dataset performance and indicating minimal sensitivity to dataset-specific biases. Logistic regression also performs well in this direction, likely due to the more constrained variability in ADReSS20 after using only full enhanced audio and unified ASR processing. In contrast, XGBoost and SVC show modest performance drops, reflecting their greater sensitivity to covariate shifts in feature distributions.

Across both validation directions, the neural networks classifier consistently delivers high and stable performance, highlighting its ability to exploit the complementary structure of concatenated acoustic and linguistic embeddings and to learn invariant representations across datasets. The symmetry of the results in Tables \ref{gen_table1} and \ref{gen_table2} indicates that the proposed multimodal representations are not overfitted to a specific dataset, preprocessing pipeline, or speaker cohort.

These cross-dataset experiments provide strong evidence that the proposed multimodal LLM-based framework generalizes well under realistic dataset shifts. By jointly modeling how speech is produced and what is being said, and by leveraging expressive multimodal representations with a suitable classifier, the proposed method achieves robust and transferable performance, an essential requirement for real-world clinical deployment of speech-based cognitive impairment screening systems.

\begin{table}[]
\caption{Generalization study wherein our model was trained on ADReSS20 data and tested on ADReSSo21 data. }
\begin{center}
\begin{tabular}{c|c|c|c|c}
\hline
\textbf{Classifier}          & \textbf{Accuracy} & \textbf{Precision} & \textbf{Recall} & \textbf{F1} \\ \hline
Neural Networks     & 91.5\%            & 93.9\%             & 88.6\%          & 91.2\%      \\ \hline
SVC                & 66.2\%            & 72.0\%             & 51.4\%          & 60.0\%      \\ \hline
XGBoost             & 88.7\%            & 90.9\%             & 85.7\%          & 88.2\%      \\ \hline
Logistic Regression & 74.6\%            & 75.8\%             & 71.4\%          & 73.5\%      \\ \hline
\end{tabular}
\end{center}
\label{gen_table1}
\end{table}

\begin{table}[]
\caption{Generalization study wherein our model was trained on ADReSSo21 data and tested on ADReSS20 data. }
\begin{center}
\begin{tabular}{c|c|c|c|c}
\hline
\textbf{Classifier}          & \textbf{Accuracy} & \textbf{Precision} & \textbf{Recall} & \textbf{F1} \\ \hline
Neural Networks     & 91.7\%            & 91.7\%             & 91.7\%          & 91.7\%      \\ \hline
SVC                & 85.4\%            & 81.5\%             & 91.7\%          & 86.3\%      \\ \hline
XGBoost             & 83.3\%            & 80.8\%             & 87.5\%          & 84.0\%      \\ \hline
Logistic Regression & 91.7\%            & 91.7\%             & 91.7\%          & 91.7\%      \\ \hline
\end{tabular}
\end{center}
\label{gen_table2}
\end{table}

\section{Conclusions and Future Work}

This work presents a generalizable, multimodal, and privacy-preserving framework for CI detection from spontaneous speech by jointly leveraging AudioLLMs and large-scale text-based LLMs. By integrating complementary acoustic and linguistic representations, the proposed approach enables more robust and informative modeling of cognitive decline. Extensive experiments conducted on the ADReSS20 and ADReSSo21 benchmark datasets demonstrate that the proposed multimodal framework consistently outperforms single-modality approaches as well as existing state-of-the-art methods. In particular, the multimodal configuration combining Qwen-based audio and text embeddings with a neural networks classifier achieves a peak classification accuracy of 92.4\%, representing a substantial improvement over prior LLM-based and traditional baselines.

Beyond overall accuracy, MMSE-stratified evaluations further highlight the effectiveness of the proposed method. These analyses show improved discriminative capability across cognitive severity levels, with particularly strong performance in identifying individuals at early stages of cognitive impairment. This capability is clinically important, as early detection enables timely intervention and more effective disease management.

A detailed comparison of classification models reveals that neural networks consistently outperform linear and tree-based classifiers when applied to high-dimensional multimodal embeddings. This observation is consistent with the limitations of logistic regression and gradient-boosted trees in modeling complex cross-modal feature interactions. Moreover, cross-dataset experiments demonstrate that the proposed framework generalizes well under realistic dataset shifts, including differences in speaker cohorts, recording conditions, and preprocessing pipelines. Such robustness is essential for real-world clinical deployment, where data heterogeneity is inevitable. Importantly, the exclusive use of open-source, locally deployable models ensures compliance with data privacy regulations and makes the framework suitable for deployment in privacy-sensitive healthcare environments.

While the proposed framework demonstrates strong performance and robustness, several directions remain for future research. First, extending the framework to support multi-class cognitive staging (e.g., CN, mild CI, and CI) and continuous MMSE score prediction. These extensions would allow more detailed clinical assessment and support tracking of cognitive changes over time. Second, expanding evaluation to multilingual and cross-linguistic settings is critical for broader applicability, given the language-dependent nature of linguistic features. Validation on additional datasets and languages will further strengthen evidence of generalizability. Finally, future work will focus on improving model interpretability by incorporating explainable AI techniques, enabling clinicians to better understand which acoustic and linguistic markers drive model decisions. Such transparency is essential for building clinical trust and facilitating adoption in real-world healthcare settings.

\section*{Competing Interests} 
The authors declare no competing interests.
\section*{Funding} 
This research was supported by funding from Saskatchewan Polytechnic, Faculty of Digital Innovation, Arts and Sciences and the 2025 Applied Research Release Time (ARRT) Fund from Saskatchewan Polytechnic.
\section*{Consent to publish} 
All authors declare consent for publication.
\section*{Data availability} 
Please request access to the datasets from https://talkbank.org/dementia/.
\section*{code availability} 
Code will be shared upon request.
\section*{Author contributions} 
Yingchao Huang: writing – review \& editing, writing – original draft, validation, visualization, formal analysis, data curation, conceptualization, methodology. Xin Wang: writing – review \& editing, validation, visualization, formal analysis, data curation, conceptualization, methodology. Yuhan Su: writing – original draft, visualization, validation, formal analysis. Wei Peng: writing – review \& editing, visualization, formal analysis. Shanshan Yao: writing – review \& editing,  visualization, validation.

\section*{Declaration of generative AI and AI-assisted technologies in the manuscript preparation process}

During the preparation of this work, the author(s) used Grammarly in order to check grammar and refine languages. After using this tool/service, the author(s) reviewed and edited the content as needed and take(s) full responsibility for the content of the published article.

\section*{Acknowledgements}
The data used in this study were obtained from the Alzheimer's Disease ADReSS20 and ADReSSo2021 databases. We gratefully acknowledge access to these valuable open-source datasets and the support provided by Saskatchewan Polytechnic. Y.Huang and X.Wang contributed equally to this work.

\bibliography{dementia_da}

\end{document}